\documentclass[lettersize,journal]{IEEEtran}
\usepackage{amsmath,amsfonts}
\usepackage{algorithmic}
\usepackage{algorithm}
\usepackage{array}
\usepackage[caption=false,font=normalsize,labelfont=sf,textfont=sf]{subfig}
\usepackage{textcomp}
\usepackage{stfloats}
\usepackage{url}
\usepackage{verbatim}
\usepackage{graphicx}
\usepackage{cite}
\usepackage[table,xcdraw]{xcolor}
\usepackage{multirow}
\begin{document}

\title{Interpretable cancer cell detection with phonon microscopy using multi-task conditional neural networks for inter-batch calibration}

\author{Yijie Zheng, Rafael Fuentes-Dominguez, Matt Clark, George S.D. Gordon, Fernando Perez-Cota\\
Optics and Photonics Research Group, University of Nottingham, NG7 2RD\\
george.gordon@nottingham.ac.uk\\
fernando.perez-cota@nottingham.ac.uk}
\maketitle

\begin{abstract}
Advances in artificial intelligence (AI) show great potential in revealing underlying information from phonon microscopy (high-frequency ultrasound) data to identify cancerous cells. However, this technology suffers from the `batch effect' that comes from unavoidable technical variations between each experiment, creating confounding variables that the AI model may inadvertently learn. We therefore present a multi-task conditional neural network framework to simultaneously achieve inter-batch calibration, by removing confounding variables, and accurate cell classification of time-resolved phonon-derived signals. We validate our approach by training and validating on different experimental batches, achieving a balanced precision of 89.22\% and an average cross-validated precision of 89.07\% for classifying background, healthy and cancerous regions. Classification can be performed in  $\sim$0.5 seconds with only simple prior batch information required for multiple batch corrections. Further, we extend our model to reconstruct denoised signals, enabling physical interpretation of salient features indicating disease state including sound velocity, sound attenuation and cell-adhesion to substrate.
\end{abstract}


\section{Introduction}
\IEEEPARstart Mechanical cancer information such as elasticity and stiffness are closely associated with cell characteristics, which could play a crucial role in early and accurate diagnosis in disease such as cancer\cite{zemla2018atomic,hayashi2015stiffness,grady2016cell}. Recent developments in label-free phonon microscopy \cite{perez2023classification,fuentes2023parallel,perez2019new,perez2016high,smith2015optically,prevedel2019brillouin,ballmann2015stimulated} employing time-resolved phonon-derived signals at high frequencies ($\sim$ 5 GHz) have demonstrated its ability to image high-resolution ($\sim$ 490–740 nm axial resolution) fixed and living cells and obtain their corresponding elastic properties with high fidelity \cite{hayashi2015stiffness,rice2017matrix}. 

In conventional low-frequency medical ultrasound, AI has been demonstrated to reveal hidden information and extract discriminative features from ultrasound signals/images without time-consuming manual inspection, providing improved quantitative results to enhance early detection and treatment of cancer \cite{gao2019new,geras2019artificial,fujioka2020utility,zahoor2020breast}. Similar for high-frequency ultrasound in phonon microscopy, a previous study has shown convolutional neural networks (CNN) are capable of learning signal features within one experiment (i.e. intra-batch) and performing 93\% classification between cancerous and normal cells \cite{perez2023classification}. 

However, a type of ‘batch effect’ problem arises when classifying signals from different experiments within the same dataset making inter-batch classification challenging. This effect can be attributed to various technical factors, including the experimental dates, temperature, humidity and also the nanofabricated tolerances. For practical reasons, cancer cells and healthy cells are measured in separate experiments, and so the combined effect of these technical factors is to create a confounding variable that is correlated with the disease status of the cells. This, in turn, causes AI models (e.g. neural networks) to overfit and prevents them from learning true features that distinguish between cancerous and normal cells.

Several methods have been developed to address the batch effect problem. In the field of microarray gene expression data, ComBat \cite{johnson2007adjusting} and limma \cite{smyth2003normalization} have been employed in normalizing batch-related information in scRNA-seq datasets but are not compatible with data that shows random variations, such as uptake rates for vaccinations. Other conventional pre-processing methods such as nested mixed effect models \cite{tung2017batch,bates2010lme4} modeled the shared random effect from each batch to remove the batch effects. More generalizable models such as mutual nearest neighbors (MNNs) \cite{haghverdi2018batch} and canonical correlation analysis (CCA) \cite{hardoon2004canonical} have been proposed to establish connections between two datasets, with fastMNN \cite{lun2019further}, Scanorama \cite{hie2019efficient}, BBKNN \cite{polanski2020bbknn} and Seurat 3 \cite{stuart2019comprehensive} explored as dimensionality reduction methods. However, These methods calibrate a single batch of data at a time, aligning it with the corresponding reference batched data, which can be computationally demanding, particularly when dealing with multiple batches of data. To incorporate various batches of data, scVI \cite{lopez2018deep} and BERMUDA \cite{wang2019bermuda} explicitly analyze cells from all batches simultaneously and jointly remove the batch effect. Nevertheless, these approaches require prior processing of the batch information, limiting their application to samples where batch metadata is not available a priori.

Neural networks (NN) have been applied to batch alignment problems as an alternative to traditional statistical methods. The DESC technique \cite{li2020deep} aims to iteratively eliminate the batch effect by learning the cluster-specific gene expression representations. This is an unsupervised learning model which makes it of low scalability and accessibility to effectiveness evaluation. Supervised learning models such as scGen \cite{lotfollahi2019scgen} use variational autoencoders that are capable of calibrating inter-batch information across scRNA-seq datasets, but its adaptability to other types of datasets has not been investigated. More recently, scGCN \cite{song2021scgcn} used a graph neural network-based model, which has been demonstrated to remove the batch effect for single-cell datasets and transfer labels across studies.  

Nevertheless, raw time-resolved phonon-derived signal traces contain significantly more information than a single value for stiffness or elasticity and represent depth variations in sound velocity, elasticity, absorption, etc. Existing methods to remove the `batch effect' of such time-series data present unique challenges. Specifically, for high-frequency time-resolved phonon-derived signals, the batch effect is not random but rather systematically associated with the experimental procedures. This means conventional methods, such as mixed-effects models, which are typically designed to address random batch effects that are assumed to be independent of the specific experimental conditions, may fail to appropriately capture and correct for the non-random correlations. Further, time-series data is considered as structured data with high temporal dependencies, whereas conventional statistical and AI methods are expected to handle observation-independent data, which is difficult to align with the sequential nature of time-series data. Finally, current methods have not been extensively employed in real-world applications within the domain of data classification.

In this paper, we present a new method that effectively learns the features of time-resolved phonon-derived signals and also conditions the batch features to simultaneously produce reliable classification results and inter-batch calibration. This concept is derived from a theoretical framework of causal inference \cite{pearl2009causal}, where in our scenario, batch features are the common causal influence of cell features and cell classification results. Our new method aims to remove the confounding.

Specifically, we present a multi-task conditional neural network framework that generates a batch-corrected latent space feature vector. We first demonstrate our framework enables a more robust representative model to calibrate the inter-batch effect across different experiments with a 20\% increase in classification precision over the baseline model. We then validate our framework by classifying inter-batch cell signals between cancerous, normal cells and background with an average precision of 89.22\% and cross-validated precision of 89.07\%. Finally, we explore the interpretability of the latent space feature vector by reconstructing the original signal from feature vectors at various conditions. We highlighted several advantages of this neural network-based framework: Firstly, it only requires $\sim$ 0.5 s to achieve both the inter-batch calibration and classification with high adaptability to multiple batches. Second, only very limited prior batch information (i.e. the date) is required to train this framework. Third, the latent space feature vector enables the reconstruction of de-noised signals, facilitating physical interpretation of the model and application of explainable AI techniques.

\section{METHODS}
\subsection{Batched data information}
The raw time-resolved phonon-derived signals (i.e. main dataset) presented in this paper was collected by phonon microscopy \cite{perez2023classification}, containing 300k time-series signal from 80 different breast cells. To perform the inter-batch calibration and classification, we defined 8 batches of data based on the experiment days and table \ref{tb:info} shows the detailed information of each batch. Specifically, there are 10 cells (37.21k individual signal input) in each batch, where each batch was performed on different, discontinuous days across several months. Normal cells (MCF 10a cell type) were measured and recorded with odd Batch IDs while cancerous cells (MDA-MB-231 cell type) were collected and recorded with even Batch IDs respectively. The baseline model introduced in \cite{perez2023classification} performing intra-batch classification was trained on 60\% of data in each batch, which was confounded by batch features. To establish that the batch effect is successfully corrected, we defined Batch 1,2, 5 and 6 as the training and validation set with a ratio of 4:1, and Batch 3, 4, 7 and 8 as the test set. The validation set is expected to provide unbiased evaluations and stopping criteria and the test set aims to examine the generalization performance of the model on unseen data. 
\begin{table}[!htbp]
\centering
\caption{Raw phonon-derived data information presented in each batch}
\label{tb:info}
\scalebox{0.85}{
\tabcolsep=0.065cm
\begin{tabular}{|c|c|c|c|c|cc|}
\hline
 &  &  &  &  & \multicolumn{2}{c|}{\textbf{Data split for classification}} \\ \cline{6-7} 
\multirow{-2}{*}{\textbf{\begin{tabular}[c]{@{}c@{}}Batch\\ ID\end{tabular}}} & \multirow{-2}{*}{\textbf{\begin{tabular}[c]{@{}c@{}}Experiment\\  day\end{tabular}}} & \multirow{-2}{*}{\textbf{Cell type}} & \multirow{-2}{*}{\textbf{\# cell}} & \multirow{-2}{*}{\textbf{\begin{tabular}[c]{@{}c@{}}\# Individual \\ signal input\end{tabular}}} & \multicolumn{1}{c|}{\textbf{Intra-batch}} & \textbf{Inter-batch} \\ \hline
 &  &  &  &  & \multicolumn{1}{c|}{\cellcolor[HTML]{7BD5EF}\textit{Train / Val}} & \cellcolor[HTML]{7BD5EF} \\ \cline{6-6}
\multirow{-2}{*}{1} & \multirow{-2}{*}{Day 1} & \multirow{-2}{*}{\begin{tabular}[c]{@{}c@{}}Normal \\ (MCF 10a)\end{tabular}} & \multirow{-2}{*}{10} & \multirow{-2}{*}{37.21 k} & \multicolumn{1}{c|}{\cellcolor[HTML]{A6DCB8}\textit{Test}} & \multirow{-2}{*}{\cellcolor[HTML]{7BD5EF}\textit{Train / Val}} \\ \hline
 &  &  &  &  & \multicolumn{1}{c|}{\cellcolor[HTML]{7BD5EF}\textit{Train / Val}} & \cellcolor[HTML]{7BD5EF} \\ \cline{6-6}
\multirow{-2}{*}{2} & \multirow{-2}{*}{Day 2} & \multirow{-2}{*}{\begin{tabular}[c]{@{}c@{}}Cancer \\ (MDA-MB-231)\end{tabular}} & \multirow{-2}{*}{10} & \multirow{-2}{*}{37.21 k} & \multicolumn{1}{c|}{\cellcolor[HTML]{A6DCB8}\textit{Test}} & \multirow{-2}{*}{\cellcolor[HTML]{7BD5EF}\textit{Train / Val}} \\ \hline
 &  &  &  &  & \multicolumn{1}{c|}{\cellcolor[HTML]{7BD5EF}\textit{Train / Val}} & \cellcolor[HTML]{A6DCB8} \\ \cline{6-6}
\multirow{-2}{*}{3} & \multirow{-2}{*}{Day 3} & \multirow{-2}{*}{\begin{tabular}[c]{@{}c@{}}Normal \\ (MCF 10a)\end{tabular}} & \multirow{-2}{*}{10} & \multirow{-2}{*}{37.21 k} & \multicolumn{1}{c|}{\cellcolor[HTML]{A6DCB8}\textit{Test}} & \multirow{-2}{*}{\cellcolor[HTML]{A6DCB8}\textit{Test}} \\ \hline
 &  &  &  &  & \multicolumn{1}{c|}{\cellcolor[HTML]{7BD5EF}\textit{Train / Val}} & \cellcolor[HTML]{A6DCB8} \\ \cline{6-6}
\multirow{-2}{*}{4} & \multirow{-2}{*}{Day 4} & \multirow{-2}{*}{\begin{tabular}[c]{@{}c@{}}Cancer \\ (MDA-MB-231)\end{tabular}} & \multirow{-2}{*}{10} & \multirow{-2}{*}{37.21 k} & \multicolumn{1}{c|}{\cellcolor[HTML]{A6DCB8}\textit{Test}} & \multirow{-2}{*}{\cellcolor[HTML]{A6DCB8}\textit{Test}} \\ \hline
 &  &  &  &  & \multicolumn{1}{c|}{\cellcolor[HTML]{7BD5EF}\textit{Train / Val}} & \cellcolor[HTML]{7BD5EF} \\ \cline{6-6}
\multirow{-2}{*}{5} & \multirow{-2}{*}{Day 5} & \multirow{-2}{*}{\begin{tabular}[c]{@{}c@{}}Normal \\ (MCF 10a)\end{tabular}} & \multirow{-2}{*}{10} & \multirow{-2}{*}{37.21 k} & \multicolumn{1}{c|}{\cellcolor[HTML]{A6DCB8}\textit{Test}} & \multirow{-2}{*}{\cellcolor[HTML]{7BD5EF}\textit{Train / Val}} \\ \hline
 &  &  &  &  & \multicolumn{1}{c|}{\cellcolor[HTML]{7BD5EF}\textit{Train / Val}} & \cellcolor[HTML]{7BD5EF} \\ \cline{6-6}
\multirow{-2}{*}{6} & \multirow{-2}{*}{Day 6} & \multirow{-2}{*}{\begin{tabular}[c]{@{}c@{}}Cancer \\ (MDA-MB-231)\end{tabular}} & \multirow{-2}{*}{10} & \multirow{-2}{*}{37.21 k} & \multicolumn{1}{c|}{\cellcolor[HTML]{A6DCB8}\textit{Test}} & \multirow{-2}{*}{\cellcolor[HTML]{7BD5EF}\textit{Train / Val}} \\ \hline
 &  &  &  &  & \multicolumn{1}{c|}{\cellcolor[HTML]{7BD5EF}\textit{Train / Val}} & \cellcolor[HTML]{A6DCB8} \\ \cline{6-6}
\multirow{-2}{*}{7} & \multirow{-2}{*}{Day 7} & \multirow{-2}{*}{\begin{tabular}[c]{@{}c@{}}Normal \\ (MCF 10a)\end{tabular}} & \multirow{-2}{*}{10} & \multirow{-2}{*}{37.21 k} & \multicolumn{1}{c|}{\cellcolor[HTML]{A6DCB8}\textit{Test}} & \multirow{-2}{*}{\cellcolor[HTML]{A6DCB8}\textit{Test}} \\ \hline
 &  &  &  &  & \multicolumn{1}{c|}{\cellcolor[HTML]{7BD5EF}\textit{Train / Val}} & \cellcolor[HTML]{A6DCB8} \\ \cline{6-6}
\multirow{-2}{*}{8} & \multirow{-2}{*}{Day 8} & \multirow{-2}{*}{\begin{tabular}[c]{@{}c@{}}Cancer \\ (MDA-MB-231)\end{tabular}} & \multirow{-2}{*}{10} & \multirow{-2}{*}{37.21 k} & \multicolumn{1}{c|}{\cellcolor[HTML]{A6DCB8}\textit{Test}} & \multirow{-2}{*}{\cellcolor[HTML]{A6DCB8}\textit{Test}} \\ \hline
\end{tabular}}
\end{table}

\begin{figure*}[!htbp]
    \centering
    \includegraphics[width=18cm]{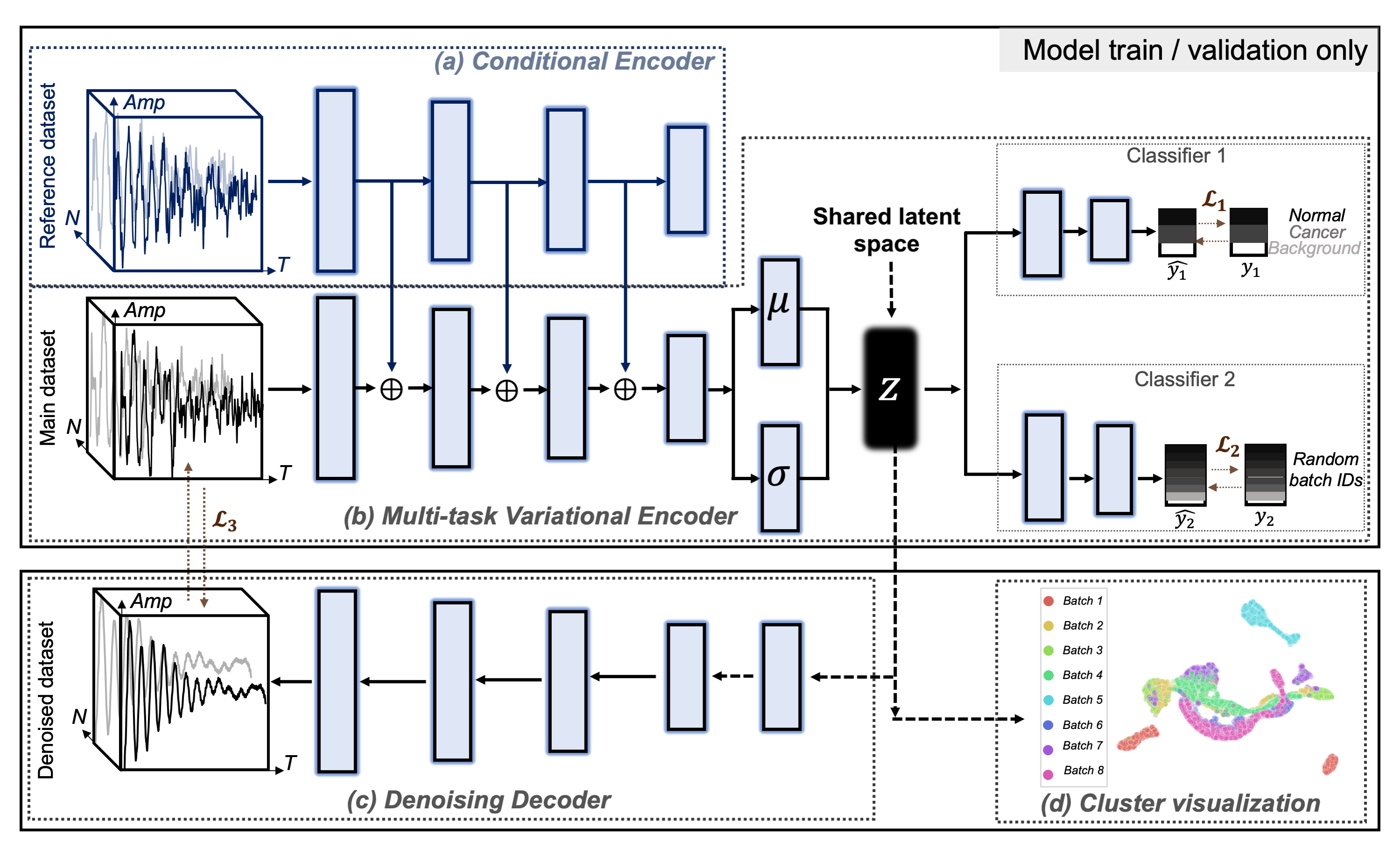}
    \caption{Multi-task conditional neural network (a) Conditional Encoder, trained with a reference dataset as a conditional reference operator. (b) Multi-task variational encoder, mapping the input data into a shared latent space with variational distribution. Also, two classifiers are trained separately where Classifier 1 aims to distinguish the normal, cancerous cells and the background, while Classifier 2 aims to prevent the model from learning the batch IDs of the data. (c) Denoising Decoder to reconstruct the input signal from the shared latent space. (d) Cluster visualization of shared latent space.}
    \label{fig:Model}
\end{figure*}

\subsection{Theoretical confounding considerations}

In theory, this `batch effect' problem can be referred to a typical example of confounding model \cite{pearl2009causal}, where batch features are the common confounding variables that associate with both cell features and the disease status of the cells. Therefore, instead of identifying the direct correlation between cell features with the disease detection, the results are confounded by the presence of batch features.

To avoid this issue, we adapt a technique that has been mathematically proven to remove confounding variables based on Bayesian network analysis by conditioning on and marginalizing over the confounding variable \cite{jager2008confounding}. First, to force marginalization over the conditioning, a multi-task output is used in which the cell classification is maximized simultaneously with minimization of the batch ID detection. This procedure randomly allocates input phonon signal traces to a Batch ID, which enforces the condition that a correctly marginalized model will be unable to predict batch ID. Second, we enforce conditioning on batch features by selecting pairs of phonon signal traces: one from a cell area and one from a randomly selected background area.

\subsection{Network architectures}
Following these principles, we construct a multi-tasking conditional neural network model shown in Figure \ref{fig:Model} that is able to simultaneously learn batch and cell features, condition the cell features on the batch features, and then learn the correct marginalization over these conditions.

Specifically, the conditional encoder (Figure \ref{fig:Model}(a)) is used to learn batch features, which are used to condition a variational encoder (VE) that learns the cell features. The conditional encoder includes four Conv2D and MaxPooling layers respectively with Leaky Relu activation function defined, exploiting efficient correlations for time series data.
Figure \ref{fig:Model}(b) shows the architecture of multi-task variational encoder with two classifiers, where the encoder which has the same structure as that of conditional encoder shown in \ref{fig:Model}(a), with a sampling layer to map the input data into a shared latent space. The rationale for using a variational model is to provide the latent space with multivariate Gaussian distributed multi-dimensional parameters ($\mu$ and $\sigma$) with higher orthogonality and generalizations compared to a simple encoder. Additionally, it enables both signal reconstruction and further generation of diverse samples through its probabilistic modelling capacity.
Classifier 1 shown in Figure \ref{fig:Model}(b) aims to classify the given signal into one of 3 classes (i.e. normal, cancer and background) representing integer values 0 to 2 as the ground truth, inclusively. In terms of Classifier 2 shown in Figure \ref{fig:Model}(b), we firstly represent each signal in integer values 1 to 8 based on its batch ID as the ground truth for test sets, and then shuffled the values to perform random batch IDs as the ground truth for train and validation sets. The shuffle process aims to remove the effect of confounding variables (i.e. batch features) by avoiding the model learning to distinguish the batch ID itself through assigning randomized ID values. This in turn, gives the model emphasis on learning the features to distinguish the normal, cancer and background of the cell.
 
The connectivity is performed by concatenating the output of each hidden layer of the conditional encoder and the corresponding output of each hidden layer of the multi-task variational encoder. This aims to provide both conditioning on and marginalizing over the confounding variable in order to remove the confounding effect from the batch features. The whole model was trained iteratively with the total loss function defined as the sum of two balanced categorical cross entropy losses performed in two classifiers respectively, giving more importance to underrepresented classes. It takes the model 145 min and 2000 epochs to converge using Tensorflow 2.0 running on an NVIDIA GeForce RTX 3090 GPU. The Adam optimizer was used with a learning rate of 0.001 in a decay rate of 1e$^{-5}$.

\subsection{Latent space analysis}
The interpretation of the framework can be represented by the latent space, which contains the latent features learned by the neural networks. We first perform the cluster visualization by plotting the Uniform Manifold Approximation and Projection (UMAP) map \cite{mcinnes2018umap}, a low-dimensional structural preserved graph, from which we expect to visualize the projected clusters from different categories as shown in Figure \ref{fig:Model}(d). 

\subsection{Signal reconstruction}
Having validated the accuracy of our classifier model, we now wish to exploit the encoder to help identify and interpret features of the input signals that denote cancer. This could help to identify physical properties indicative of cancer and thus represent a form of explainable AI.

Specifically, beginning with the trained latent space, we train a denoising decoder model as shown in Figure \ref{fig:Model} (c). This decoder model aims to reconstruct the original signal from the low-dimensional latent space that contains the features crucial for accurate calibration and classification. The model is expected to produce denoised signals that preserve the critical components contributing to the classification while filtering out unimportant and batch-specific features.

\subsection{Physical features extraction for model interpretation}
To extract physical parameters from the phonon-derived signal, we have used a simple model which considers the cell to be homogeneous. The model is based on time-resolved Brillouin scattering where the signal is of sinusoidal nature due to optical interference and its decay exponential due to sound attenuation. 

\begin{equation}
    signal =  A \cos(2\pi f z +\phi)e^{(-\alpha_0 z)}
    \label{eq:TRBS}
\end{equation}
\noindent
where $f$ is the Brillouin frequency, $\nu$ is the sound attenuation coefficient, $\phi$ is the signal phase, $A$ the initial amplitude and $z$ is the distance from the transducer.  The Brillouin frequency is then a function of the sound velocity $v$: 
\begin{equation}
    f =  2n\nu /\lambda
    \label{eq:TRBS}
\end{equation}
\noindent
where $n$ is the refractive index of the cell and $\lambda$ the optical probe wavelength.  The sound velocity is an elastic property of the material which is related to the density and longitudinal modulus of the cell. The experimental signal is then fitted to equation \ref{eq:TRBS} to the minimum possible error resulting in the estimation of the sound velocity, the sound attenuation coefficient and the signal's phase. 
\subsection{Evaluation metrics}
We define several evaluation metrics in order to quantitatively access the performance of classification using this multi-task conditional neural network, which can be visualized through a confusion matrix. Here, the confusion matrix shows the frequency map of multi-class predictions, where each cell in the matrix contains the count or frequency of instances belonging to a specific actual class and predicted class pair. For each actual class (i.e. each row), we define the frequency value of correct predictions at the $i^{th}$ class (where both the actual and predicted labels are within the same class) as $C_i$ among normal ($i=1$), cancerous ($i=2$) and background ($i=3$), and false predictions as $F_i$.

To assess the overall accuracy of our model, we employ a balanced precision metric \cite{scikit-learn}, which is well-suited for our imbalanced datasets. This metric computes the weighted average precision across the whole test set, taking into account the relative sizes of different classes:
\begin{equation}
    \begin{aligned}
        \mathrm{balanced\ precision} = &{W_1}\times\frac{C_1}{C_1+F_1}+{W_2}\times\frac{C_2}{C_2+F_2}\\
        &+{W_3}\times\frac{C_3}{C_3+F_3}
    \end{aligned}
\end{equation}
\begin{equation}
    W_i = \frac{\frac{1}{y_i}}{{\frac{1}{y_1}+\frac{1}{y_2}+\frac{1}{y_3}}}
\end{equation}
\noindent where, ${W_i}$ represents the weight assigned to the $i^{th}$ class and $y_i$ represents the number of test instances belonging to the $i^{th}$ class. This means classes with fewer instances (smaller $y_i$ values) will have a higher weight, while classes with more instances (larger $y_i$ values) will have a lower weight.

We next evaluate the model's predictive capability for individual batches and gauge any variations in its performance by calculating the average precision for each cell from different batches:
\begin{equation}
    \mathrm{Precision} = \frac{1}{N}\times \sum_{n=1}^{N}\mathrm{balanced \ precision}
\end{equation}
\noindent where, for each batch, $N$ represents the total number of time-series signals associated with each cell, which corresponds to the number of classified results per cell (3721 in this paper). Therefore, this metric contains a vector with a length of 10 for each batch where each element in the vector corresponds to the average balanced precision of a specific cell.

Additionally, we measure the sensitivity and specificity of the model to evaluate its ability to accurately differentiate between cancerous and normal cells. High sensitivity is expected to minimize false negatives and ensure the test is effective in catching cancer cases early while high specificity ensures that individuals without cancer are not subjected to unnecessary diagnostic procedures or treatments. 

\section{RESULTS}
\subsection{Validation of inter-batch calibration}
\begin{figure*}[!htbp]
    \centering
    \includegraphics[width=17cm]{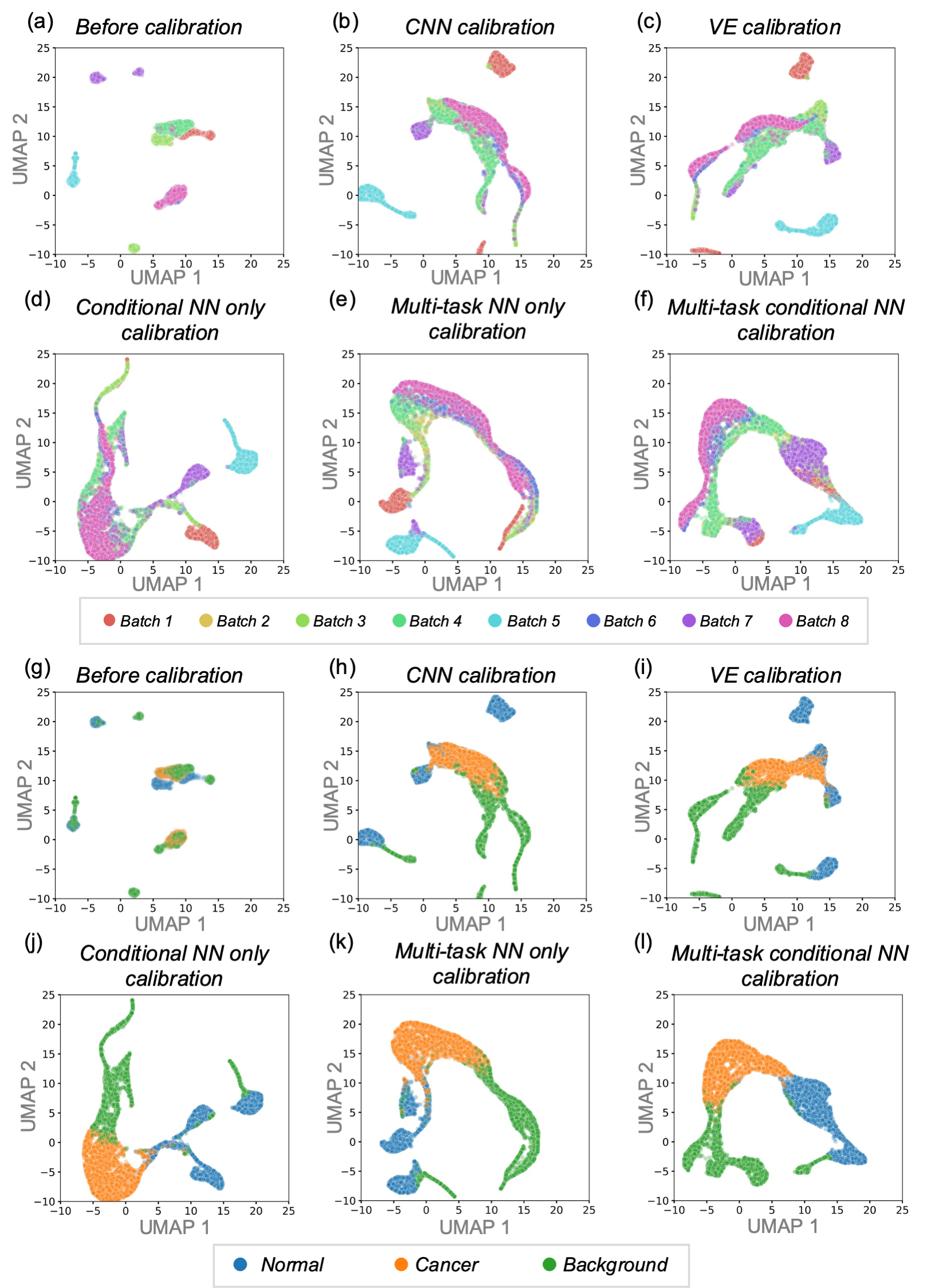}
    \caption{UMAP clustering plot (a) before and (b)-(f) after calibration with various calibration models, colored by 8 different batch IDs. Corresponding UMAP plot (g) - (l), colored by 3 different classes (i.e. normal, cancer, and background).}
    \label{fig:calibration}
\end{figure*}

We assess the effectiveness of calibrating the inter-batch effect by examining the clustering plot of the latent space, which represents the critical features utilized for classification. Figure \ref{fig:calibration}(a) shows the UMAP plot of the raw data before the calibration, and (b) to (f) show the UMAP plot of the latent space original and calibrated signal by our multi-task conditional NN, with colors representing different batch IDs. The raw data before calibration exhibit more diffused clustering, especially with Batch 5 and Batch 7 distinctly separated from the others. This explains why features obtained from specific batches for classification may not adequately represent the rest batches of data, resulting in poor generalisability for separating different classes, as shown in Figure \ref{fig:calibration}(g), where the UMAP plot is colored by different classes. In contrast, after calibration, the clustering results shown in Figure \ref{fig:calibration}(f) show that data from different batches is more contiguously distributed, with 3 classes highly separable in Figure \ref{fig:calibration}(l). This demonstrates effective batch calibration because different batches now map to the same regions in the clustering map.

We next compared our model with other neural network models, namely the baseline model using CNN presented in \cite{perez2023classification}, variational encoder (VE) \cite{lotfollahi2019scgen} commonly used for normalization and calibration, conditional NN-only model without Classifier 2, and multi-task NN only model without background reference input. Figure \ref{fig:calibration}(b)-(c) show a more condensed distribution in clustering maps for different Batch IDs compared to the result before calibration. However, Batch 1 and Batch 5 clusters are still dispersed far apart, making it challenging to separate different cell classes easily as shown in Figure \ref{fig:calibration}(h)-(i). Figure \ref{fig:calibration}(d)-(e) show a further compact distribution in clustering map by using either conditional NN only or multi-task NN only model respectively. This leads to improved separation between different cell classes, as shown in Figure \ref{fig:calibration}(j)-(k). Considering this, the combination of both models(i.e. our multi-task conditional NN) results in the most continuously distributed UMAP plot where batch IDs may not be distinctly separated, but different classes can still be distinguished effectively as shown in Figure \ref{fig:calibration}(l).

Therefore, it demonstrates that both conditioning and marginalization to control the confounding variables are theoretically necessary to remove confounding variables. However, the multi-task conditional model which enforces both conditioning and marginalization shows its optimal capability in mitigating confounding problems and achieving accurate cell classification. 
\subsection{Validation of inter-batch classification}
\begin{figure*}
    \centering
    \includegraphics[width=18cm]{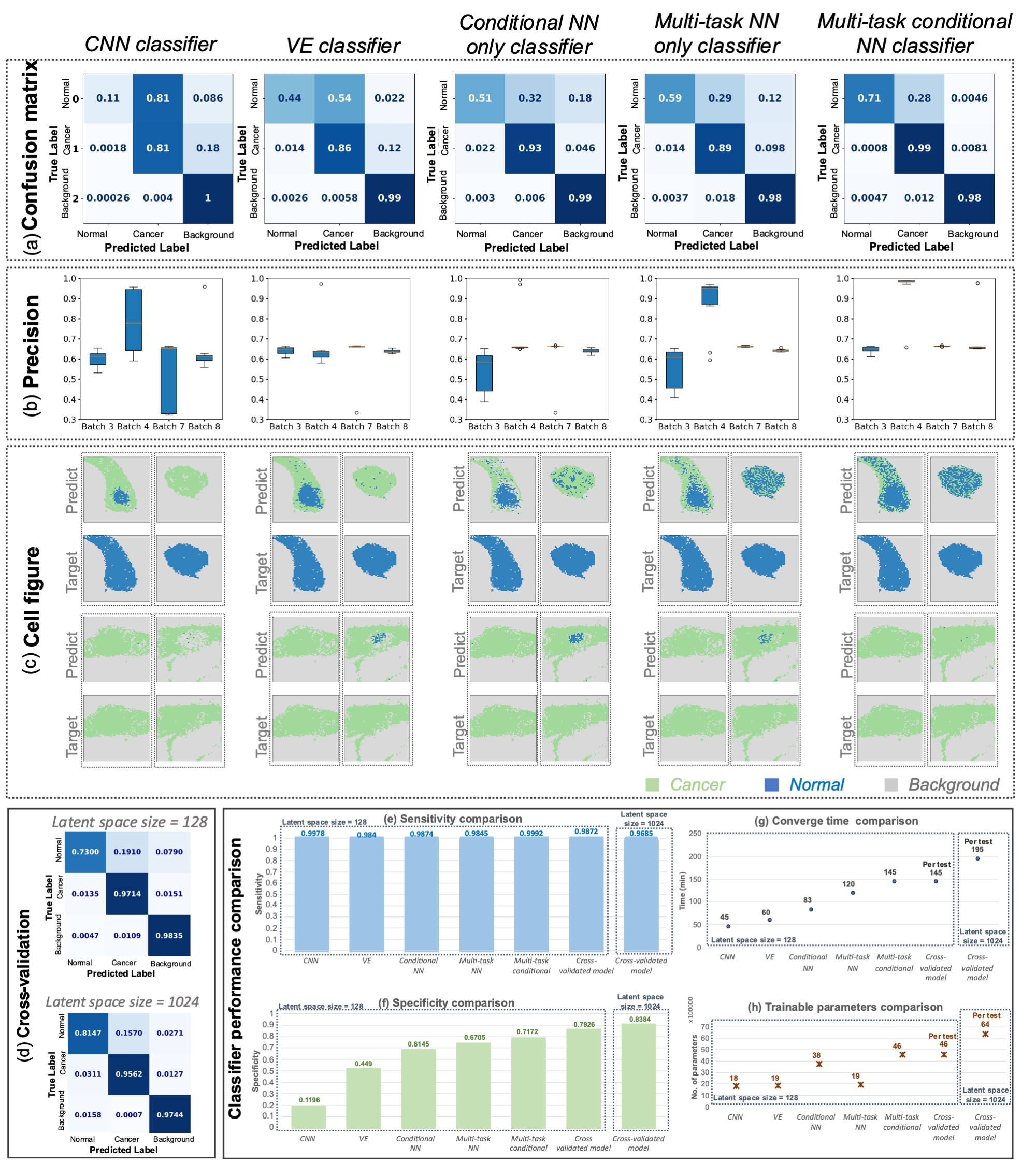}
    \caption{Inter-batch classification results of various classifiers evaluated by metrics (a) confusion matrix, (b) average precision distribution of each tested batch, (c) example cell imaging, (d) confusion matrix after 6-fold cross-validation, (e) sensitivity, (f) specificity. Computational resources comparison between various classifiers: (g) converge time and (h) Trainable parameters}
    \label{fig:classification}
\end{figure*}

We evaluate the classification performance in comparison with various models using Batch 1,2, 5 and 6 as the training sets, as shown in Figure \ref{fig:classification}. We first generate the confusion matrix shown in Figure \ref{fig:classification}(a) for different models and then calculate the balanced precision by assigning individual weights to each class with a ratio of 0.343:0.48:0.177. The baseline model demonstrates a 100\% capability in distinguishing the background class from the others. However, it exhibits limited ability to differentiate between cancerous and normal cells, as the majority (81\%) of predictions that were expected to be normal cells are instead assigned to the cancerous classes. The commonly used calibration model (i.e. variational encoder) shows improved behavior in classifying the cancerous and normal cells, achieving a balanced precision of 73.90\%, which is 13\% greater than the baseline model. Both the multi-task-only model and the conditional-only model achieve significantly higher balanced precision of 79.66\% and 80.30\%, respectively. Further, the combined model (the multi-task conditional model) achieves the highest balanced precision of 89.22\%, indicating that it correctly classifies the largest portions of cell classes. 

To compare the model robustness between batches, we next calculate the average precision for each individual cell within a batch, which means averaging across full prediction results (i.e. 3721 points) per cell within each batch. We then represent these average precision results in each batch using boxplots in Figure \ref{fig:classification}(b), where the blue box extends from the Q1 to Q3 quartile of the data distribution, with the orange line indicating the median (Q2) data value. The multi-task conditional NN classifier improves the median average precision value across the four tested batches of data, especially in Batch 4, where the median precision increases from 0.77 to 0.99. Also, it shows the lowest standard deviation of the average precision values, with less than 0.2 across all four batches, indicating the generalization and robustness ability of this model.

Additionally, we plot the predicted and target cell example images colored by class ID from four tested batches in Figure \ref{fig:classification}(c), where green and blue represent the cancerous and normal pixels respectively, with grey representing the background (water). Our multi-task conditional model visually shows the best predictions, especially for predicting normal cells, where both baseline model and VE classifier has limited capacity to distinguish the normal from cancerous classes. This further demonstrates that this model has the highest specificity of 0.7172 as shown in Figure \ref{fig:classification}(e), which means it can correctly ensure the cell is not cancer. However, all five models show a superior high sensitivity with the highest of 0.9992 in Figure \ref{fig:classification}(f) from the multi-task conditional NN model, which means the model is effective at detecting individuals who actually have the disease, minimizing the chances of false negatives.

To assess the model's generalizability, we conducted a 6-fold cross-validation, evaluating performance by assigning various data batches as training sets to prevent overfitting to specific subsets. The figure \ref{fig:classification}(d) shows the cross-validated average confusion matrix, with an average balanced precision of 89.07\% and a standard variation of 0.99\%, indicating a commendable generalization ability of our model. Additionally, enlarging the size of the latent space from 128 to 1024 resulted in an increased balanced precision of 91.09\%. However, this leads to trade-off considerations, as it leads to longer training times and higher memory usage.

We finally compared the computational resources of these models which are presented in Figure \ref{fig:classification}(g) and (h): converge and prediction time, trainable parameters comparison. Although the combinational model requires the longest training time (145 min) and the largest trainable parameters (46 million) compared to the other models, it only takes less than 0.5s to achieve the prediction, which is desirable for this task.
\subsection{Denoised signal reconstruction from latent space for explainability}
\begin{figure*}
    \centering
    \includegraphics[width=18cm]{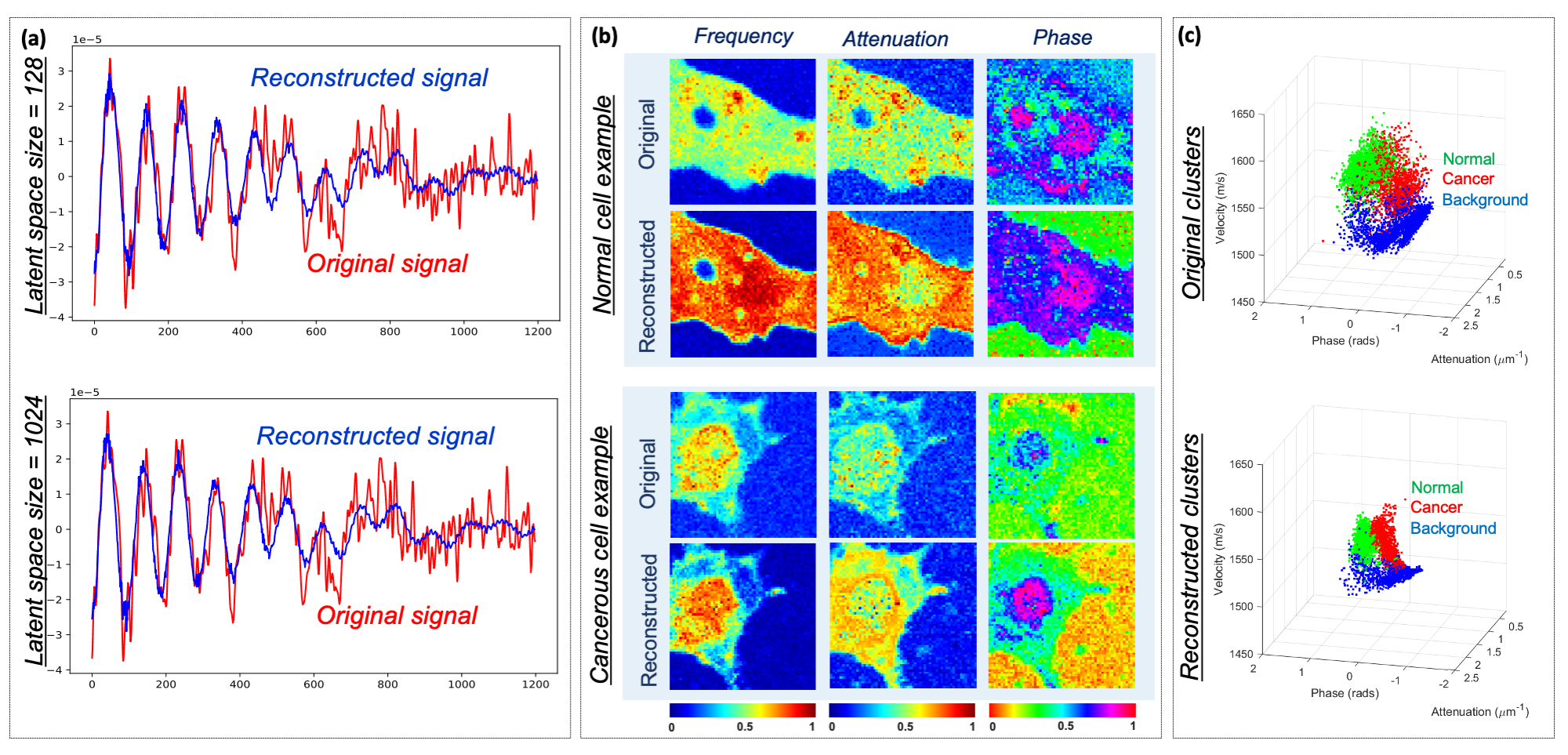}
    \caption{Explainability of latent space for signal denoising. (a) Denoised signal reconstructed from latent space with dimension 128 and 1024. (b) Feature (Frequency, Attenuation and Phase) map of normal and cancerous cell examples from the reconstructed and original signal. (c) Feature (Frequency, Attenuation and Phase) cluster from the reconstructed and original signal, where blue represents background, red represents cancer and green represents normal.}
    \label{fig:signal}
\end{figure*}

Figure \ref{fig:signal}(a) shows the denoised signal reconstructed from the decoder model, with the latent space size of 128 and 1024 respectively. In general, the larger latent space will contain more comprehensive features, which produce a slightly higher classification precision of 89.9\% compared to that of 82.2\% (Figure \ref{fig:classification}(a)) but requires twice as many trainable parameters and double the training time. Also, the reconstructed signals from both latent space sizes exhibit visually similar behavior. This indicates that the size of 128 already contains the most important features, and further increases in size may not result in significant improvements. This provides an initial interpretation of this model by filtering out the unimportant components.

The interpretability becomes more evident when employing a basic physical model introduced in \cite{perez2023classification}. The frequency of the signal correlates to the sound velocity, its decay rate to the sound attenuation coefficient, and its phase, among other factors, to the cell's adhesion characteristics to the substrate. These fitted parameters can be visualized as images to highlight important structural features of cells that may be associated with the disease. Is possible to see that the contrast before and after denoising changes (see Figure \ref{fig:signal}(b)), reveals features that were previously unclear. Specifically, the de-noised phase component provides a much clearer definition of the cell's boundary and improves contrast for spatial structures within the nucleus. Visualizing these physical parameters on a scatter plot (see Figure \ref{fig:signal}(c)), allows us to discern differences between these characteristics across different classes. Before denoising, the scatter plot appears diffuse with significant overlap between pixel class clusters.  Following denoising, the clusters are much more tightly packed with less overlap. To quantify this enhancement, we trained different conventional classification models, namely Linear, SVM, Kernel, Discriminant and Naive Bayes classifiers from physical features that have been extracted (velocity, attenuation and phase). Table \ref{tab:classification} shows the weighted precision of different models trained from 2D/3D features (where 2D includes velocity and attenuation features and 3D includes velocity, attenuation and phase features). Compared to 2D classification, where both original and reconstruction signals show relatively low precision, 3D classification shows higher precision, especially from the reconstructed signals, where the weighted precision results are over 90\% from all models (details in Supplementary Document). This means the phase feature contributes significantly to the classification task but the inherent noise of this feature in the raw data is too high to allow interpretation.

\begin{table}[!htbp]
\caption{Weighted precision of classifications different conventional classification models trained from physical features}
\label{tab:classification}
\resizebox{0.48\textwidth}{!}{%
\begin{tabular}{|c|c|ccccc|}
\hline
\rowcolor[HTML]{FFFFFF} 
\cellcolor[HTML]{FFFFFF} &
  \cellcolor[HTML]{FFFFFF} &
  \multicolumn{5}{c|}{\cellcolor[HTML]{FFFFFF}\textbf{Conventional classification models}} \\ \cline{3-7} 
\rowcolor[HTML]{FFFFFF} 
\multirow{-2}{*}{\cellcolor[HTML]{FFFFFF}\textbf{\begin{tabular}[c]{@{}c@{}}Feature \\ size\end{tabular}}} &
  \multirow{-2}{*}{\cellcolor[HTML]{FFFFFF}\textbf{Data Input}} &
  \multicolumn{1}{c|}{\cellcolor[HTML]{FFFFFF}\textit{Linear}} &
  \multicolumn{1}{c|}{\cellcolor[HTML]{FFFFFF}\textit{SVM}} &
  \multicolumn{1}{c|}{\cellcolor[HTML]{FFFFFF}\textit{Kernel}} &
  \multicolumn{1}{c|}{\cellcolor[HTML]{FFFFFF}\textit{Discriminant}} &
  \textit{Naive Bayes} \\ \hline
\rowcolor[HTML]{FFFFFF} 
\cellcolor[HTML]{FFFFFF} &
  \textbf{Original} &
  \multicolumn{1}{c|}{\cellcolor[HTML]{FFFFFF}0.5581} &
  \multicolumn{1}{c|}{\cellcolor[HTML]{FFFFFF}0.6695} &
  \multicolumn{1}{c|}{\cellcolor[HTML]{FFFFFF}0.5864} &
  \multicolumn{1}{c|}{\cellcolor[HTML]{FFFFFF}0.6595} &
  0.6827 \\ \cline{2-7} 
\rowcolor[HTML]{FFFFFF} 
\multirow{-2}{*}{\cellcolor[HTML]{FFFFFF}\textbf{2D}} &
  \textbf{Reconstructed} &
  \multicolumn{1}{c|}{\cellcolor[HTML]{FFFFFF}0.4701} &
  \multicolumn{1}{c|}{\cellcolor[HTML]{FFFFFF}0.6696} &
  \multicolumn{1}{c|}{\cellcolor[HTML]{FFFFFF}0.6293} &
  \multicolumn{1}{c|}{\cellcolor[HTML]{FFFFFF}0.6429} &
  0.6893 \\ \hline
\rowcolor[HTML]{EFEFEF} 
\cellcolor[HTML]{EFEFEF} &
  \textbf{Original} &
  \multicolumn{1}{c|}{\cellcolor[HTML]{EFEFEF}0.6826} &
  \multicolumn{1}{c|}{\cellcolor[HTML]{EFEFEF}0.7774} &
  \multicolumn{1}{c|}{\cellcolor[HTML]{EFEFEF}0.7176} &
  \multicolumn{1}{c|}{\cellcolor[HTML]{EFEFEF}0.7608} &
  0.7773 \\ \cline{2-7} 
\rowcolor[HTML]{EFEFEF} 
\multirow{-2}{*}{\cellcolor[HTML]{EFEFEF}\textbf{3D}} &
  {\color[HTML]{010066} \textbf{Reconstructed}} &
  \multicolumn{1}{c|}{\cellcolor[HTML]{EFEFEF}{\color[HTML]{010066} 0.9000}} &
  \multicolumn{1}{c|}{\cellcolor[HTML]{EFEFEF}{\color[HTML]{010066} 0.9202}} &
  \multicolumn{1}{c|}{\cellcolor[HTML]{EFEFEF}{\color[HTML]{010066} 0.9335}} &
  \multicolumn{1}{c|}{\cellcolor[HTML]{EFEFEF}{\color[HTML]{010066} 0.9069}} &
  {\color[HTML]{010066} 0.9035} \\ \hline
\end{tabular}%
}
\end{table}
\section{DISCUSSION}
We addressed the `batch effect' problem that arises during the inter-batch classification of time-resolved phonon-derived signals, where the baseline model and AI-based Variational Encoder (VE) models have shown limited capacity to distinguish the cancerous and normal cells due to confounding factors. To overcome this, we developed a neural network-based approach: multi-task conditional NN, incorporating a conditional reference operator at the input and two separate classifiers trained simultaneously to learn the distinguishing features of healthy and cancer while being unable to determine batch ID. Using this approach, we successfully achieved removing batch confounding variables by adding conditioning and marginalization, in accordance with established results from Bayesian network theory.  Further, we show that having removed this confounding effect we can reconstruct denoised versions of the original signals with interpretable physical features useful for analysis of tissue disease states.

There are several major advantages to our neural network approach over existing approaches. First, the prediction time is very fast (typically $\sim$0.5s) even when dealing with multiple batches of data. Also, the use of background conditioning means it only needs to be trained once per application, regardless of changes in experimental equipment or days. This further makes it a feasible approach for future data expansion and other applications with larger datasets. Second, the only information required for training the model is a batch identifier that can be used to achieve randomization, though more advanced models could potentially incorporate other experimental variables to examine the influence of factors such as temperature. This is important in practical clinical scenarios where recording batch information can be challenging and the limited information will lead to poor calibration and classification performance. Finally, the model generates an interpretable low-dimensional shared latent space which can be represented in cluster visualization and used for effective signal denoising. Further analysis can explore the correlation between denoised signal characteristics (e.g. its frequency, attenuation, phase, etc.) and different classes of cells. In particular, we find that the phase information of our time-domain signals differs significantly between healthy and cancer cells.  This difference is not visible in the raw data but becomes clear in reconstructed denoised versions.  This may imply that cell adhesion to the substrate, one factor that influences such a phase shift, differs significantly between healthy and cancer cells. There have been a small number of studies \cite{varani1987tumor} previously that have suggested this but our analysis suggests this could be an important feature that is detectable using our techniques. Our AI approach enables this to be used as an explainable future that may contribute to future applications in biomedical science or medical diagnostics.

Several improvements should be considered when further exploring this approach. Firstly, the dataset used in this paper is collected over 8 experimental days, with 8 batches of data in total, containing a relatively limited diversity of the samples with a total number of 80 cells ($\sim$ 300,000 input time traces). Although our approach has been validated with high calibration and classification accuracy, assessing the model's robustness and generalizability requires testing it with a larger and more diverse dataset. This can be achieved by incorporating additional data collected on various experimental days, under different settings, and involving diverse cell types. Secondly, to implement our approach, it is necessary to know one pixel of the background for each cell as the reference operator before training the multi-task conditional model. This can be solved by developing a reliable experimental background extractor that can consistently generate the time trace of one pixel of background every time during the measurement.

This AI-driven approach shows great promise for future explainable signal analysis in the area of medical healthcare. For example, It could be applied to explore drugs that modulate elasticity and examine their effects on key physical parameters over time. Also, it could facilitate a better understanding of complex relationships within biological systems (e.g. substrate adhesion), enabling the prediction of dynamic changes over time by employing this time-series forecasting AI model, paving the way for innovative treatments and therapies.

\bibliographystyle{unsrt}  
\bibliography{references}

\begin{thebibliography}{10}

\bibitem{zemla2018atomic}
Joanna Zem{\l}a, Joanna Danilkiewicz, Barbara Orzechowska, Joanna Pabijan, Sara Seweryn, and Ma{\l}gorzata Lekka.
\newblock Atomic force microscopy as a tool for assessing the cellular elasticity and adhesiveness to identify cancer cells and tissues.
\newblock In {\em Seminars in cell \& developmental biology}, volume~73, pages 115--124. Elsevier, 2018.

\bibitem{hayashi2015stiffness}
Kozaburo Hayashi and Mayumi Iwata.
\newblock Stiffness of cancer cells measured with an afm indentation method.
\newblock {\em Journal of the mechanical behavior of biomedical materials}, 49:105--111, 2015.

\bibitem{grady2016cell}
Martha~E Grady, Russell~J Composto, and David~M Eckmann.
\newblock Cell elasticity with altered cytoskeletal architectures across multiple cell types.
\newblock {\em Journal of the mechanical behavior of biomedical materials}, 61:197--207, 2016.

\bibitem{perez2023classification}
Fernando P{\'e}rez-Cota, Giovanna Mart{\'\i}nez-Arellano, Salvatore La~Cavera~III, William Hardiman, Luke Thornton, Rafael Fuentes-Dom{\'\i}nguez, Richard~J Smith, Alan McIntyre, and Matt Clark.
\newblock Classification of cancer cells at the sub-cellular level by phonon microscopy using deep learning.
\newblock {\em Scientific Reports}, 13(1):16228, 2023.

\bibitem{fuentes2023parallel}
Rafael Fuentes-Dom{\'\i}nguez, Mengting Yao, William Hardiman, Kerry Setchfield, Fernando P{\'e}rez-Cota, Richard~J Smith, Matt Clark, et~al.
\newblock Parallel imaging with phonon microscopy using a multi-core fibre bundle detection.
\newblock {\em Photoacoustics}, 31:100493, 2023.

\bibitem{perez2019new}
Fernando P{\'e}rez-Cota, Richard~J Smith, Hany~M Elsheikha, and Matt Clark.
\newblock New insights into the mechanical properties of acanthamoeba castellanii cysts as revealed by phonon microscopy.
\newblock {\em Biomedical optics express}, 10(5):2399--2408, 2019.

\bibitem{perez2016high}
Fernando P{\'e}rez-Cota, Richard~J Smith, Emilia Moradi, Leonel Marques, Kevin~F Webb, and Matt Clark.
\newblock High resolution 3d imaging of living cells with sub-optical wavelength phonons.
\newblock {\em Scientific Reports}, 6(1):39326, 2016.

\bibitem{smith2015optically}
Richard~J Smith, Fernando~Perez Cota, Leonel Marques, Xuesheng Chen, Ahmet Arca, Kevin Webb, Jonathon Aylott, Micheal~G Somekh, and Matt Clark.
\newblock Optically excited nanoscale ultrasonic transducers.
\newblock {\em The Journal of the Acoustical Society of America}, 137(1):219--227, 2015.

\bibitem{prevedel2019brillouin}
Robert Prevedel, Alba Diz-Mu{\~n}oz, Giancarlo Ruocco, and Giuseppe Antonacci.
\newblock Brillouin microscopy: an emerging tool for mechanobiology.
\newblock {\em Nature methods}, 16(10):969--977, 2019.

\bibitem{ballmann2015stimulated}
Charles~W Ballmann, Jonathan~V Thompson, Andrew~J Traverso, Zhaokai Meng, Marlan~O Scully, and Vladislav~V Yakovlev.
\newblock Stimulated brillouin scattering microscopic imaging.
\newblock {\em Scientific Reports}, 5(1):18139, 2015.

\bibitem{rice2017matrix}
AJ~Rice, E~Cortes, D~Lachowski, BCH Cheung, SA~Karim, JP~Morton, and A~Del Rio~Hernandez.
\newblock Matrix stiffness induces epithelial--mesenchymal transition and promotes chemoresistance in pancreatic cancer cells.
\newblock {\em Oncogenesis}, 6(7):e352--e352, 2017.

\bibitem{gao2019new}
Yiming Gao, Krzysztof~J Geras, Alana~A Lewin, and Linda Moy.
\newblock New frontiers: an update on computer-aided diagnosis for breast imaging in the age of artificial intelligence.
\newblock {\em AJR. American journal of roentgenology}, 212(2):300, 2019.

\bibitem{geras2019artificial}
Krzysztof~J Geras, Ritse~M Mann, and Linda Moy.
\newblock Artificial intelligence for mammography and digital breast tomosynthesis: current concepts and future perspectives.
\newblock {\em Radiology}, 293(2):246--259, 2019.

\bibitem{fujioka2020utility}
Tomoyuki Fujioka, Mio Mori, Kazunori Kubota, Jun Oyama, Emi Yamaga, Yuka Yashima, Leona Katsuta, Kyoko Nomura, Miyako Nara, Goshi Oda, et~al.
\newblock The utility of deep learning in breast ultrasonic imaging: a review.
\newblock {\em Diagnostics}, 10(12):1055, 2020.

\bibitem{zahoor2020breast}
Saliha Zahoor, Ikram~U Lali, Muhammad~Attique Khan, Kashif Javed, and Waqar Mehmood.
\newblock Breast cancer detection and classification using traditional computer vision techniques: a comprehensive review.
\newblock {\em Current medical imaging}, 16(10):1187--1200, 2020.

\bibitem{johnson2007adjusting}
W~Evan Johnson, Cheng Li, and Ariel Rabinovic.
\newblock Adjusting batch effects in microarray expression data using empirical bayes methods.
\newblock {\em Biostatistics}, 8(1):118--127, 2007.

\bibitem{smyth2003normalization}
Gordon~K Smyth and Terry Speed.
\newblock Normalization of cdna microarray data.
\newblock {\em Methods}, 31(4):265--273, 2003.

\bibitem{tung2017batch}
Po-Yuan Tung, John~D Blischak, Chiaowen~Joyce Hsiao, David~A Knowles, Jonathan~E Burnett, Jonathan~K Pritchard, and Yoav Gilad.
\newblock Batch effects and the effective design of single-cell gene expression studies.
\newblock {\em Scientific reports}, 7(1):39921, 2017.

\bibitem{bates2010lme4}
Douglas~M Bates.
\newblock lme4: Mixed-effects modeling with r, 2010.

\bibitem{haghverdi2018batch}
Laleh Haghverdi, Aaron~TL Lun, Michael~D Morgan, and John~C Marioni.
\newblock Batch effects in single-cell rna-sequencing data are corrected by matching mutual nearest neighbors.
\newblock {\em Nature biotechnology}, 36(5):421--427, 2018.

\bibitem{hardoon2004canonical}
David~R Hardoon, Sandor Szedmak, and John Shawe-Taylor.
\newblock Canonical correlation analysis: An overview with application to learning methods.
\newblock {\em Neural computation}, 16(12):2639--2664, 2004.

\bibitem{lun2019further}
A~Lun.
\newblock Further mnn algorithm development.
\newblock {\em GitHub repository}, 2019.

\bibitem{hie2019efficient}
Brian Hie, Bryan Bryson, and Bonnie Berger.
\newblock Efficient integration of heterogeneous single-cell transcriptomes using scanorama.
\newblock {\em Nature biotechnology}, 37(6):685--691, 2019.

\bibitem{polanski2020bbknn}
Krzysztof Pola{\'n}ski, Matthew~D Young, Zhichao Miao, Kerstin~B Meyer, Sarah~A Teichmann, and Jong-Eun Park.
\newblock Bbknn: fast batch alignment of single cell transcriptomes.
\newblock {\em Bioinformatics}, 36(3):964--965, 2020.

\bibitem{stuart2019comprehensive}
Tim Stuart, Andrew Butler, Paul Hoffman, Christoph Hafemeister, Efthymia Papalexi, William~M Mauck, Yuhan Hao, Marlon Stoeckius, Peter Smibert, and Rahul Satija.
\newblock Comprehensive integration of single-cell data.
\newblock {\em Cell}, 177(7):1888--1902, 2019.

\bibitem{lopez2018deep}
Romain Lopez, Jeffrey Regier, Michael~B Cole, Michael~I Jordan, and Nir Yosef.
\newblock Deep generative modeling for single-cell transcriptomics.
\newblock {\em Nature methods}, 15(12):1053--1058, 2018.

\bibitem{wang2019bermuda}
Tongxin Wang, Travis~S Johnson, Wei Shao, Zixiao Lu, Bryan~R Helm, Jie Zhang, and Kun Huang.
\newblock Bermuda: a novel deep transfer learning method for single-cell rna sequencing batch correction reveals hidden high-resolution cellular subtypes.
\newblock {\em Genome biology}, 20(1):1--15, 2019.

\bibitem{li2020deep}
Xiangjie Li, Kui Wang, Yafei Lyu, Huize Pan, Jingxiao Zhang, Dwight Stambolian, Katalin Susztak, Muredach~P Reilly, Gang Hu, and Mingyao Li.
\newblock Deep learning enables accurate clustering with batch effect removal in single-cell rna-seq analysis.
\newblock {\em Nature communications}, 11(1):2338, 2020.

\bibitem{lotfollahi2019scgen}
Mohammad Lotfollahi, F~Alexander Wolf, and Fabian~J Theis.
\newblock scgen predicts single-cell perturbation responses.
\newblock {\em Nature methods}, 16(8):715--721, 2019.

\bibitem{song2021scgcn}
Qianqian Song, Jing Su, and Wei Zhang.
\newblock scgcn is a graph convolutional networks algorithm for knowledge transfer in single cell omics.
\newblock {\em Nature communications}, 12(1):3826, 2021.

\bibitem{pearl2009causal}
Judea Pearl.
\newblock Causal inference in statistics: An overview.
\newblock 2009.

\bibitem{jager2008confounding}
KJ~Jager, C~Zoccali, A~Macleod, and FW~Dekker.
\newblock Confounding: what it is and how to deal with it.
\newblock {\em Kidney international}, 73(3):256--260, 2008.

\bibitem{mcinnes2018umap}
Leland McInnes, John Healy, and James Melville.
\newblock Umap: Uniform manifold approximation and projection for dimension reduction.
\newblock {\em arXiv preprint arXiv:1802.03426}, 2018.

\bibitem{scikit-learn}
F.~Pedregosa, G.~Varoquaux, A.~Gramfort, V.~Michel, B.~Thirion, O.~Grisel, M.~Blondel, P.~Prettenhofer, R.~Weiss, V.~Dubourg, J.~Vanderplas, A.~Passos, D.~Cournapeau, M.~Brucher, M.~Perrot, and E.~Duchesnay.
\newblock Scikit-learn: Machine learning in {P}ython.
\newblock {\em Journal of Machine Learning Research}, 12:2825--2830, 2011.

\bibitem{varani1987tumor}
James Varani, Paul~E McKeever, Vishva Dixit, Thomas~E Carey, and Suzanne~EG Flgiel.
\newblock Tumor type-specific differences in cell-substrate adhesion among human tumor cell lines.
\newblock {\em International journal of cancer}, 39(3):397--403, 1987.

\end{thebibliography}
\vfill

\end{document}